\documentclass[useAMS, usenatbib, amssymb]{mn2e}
\usepackage{psfig}
\usepackage{graphicx}
\usepackage{epsf}
\usepackage{bm}
\usepackage{lscape}

\title [Open cluster NGC 6866]
{A comprehensive study of the open cluster NGC 6866}
\author[Bostanc\i~et~al.]
       {Z.~F. Bostanc\i $^{1}$\thanks{E-mail: funda.bostanci@istanbul.edu.tr},
T. Ak$^{1}$, T. Yontan$^{2}$, S. Bilir$^{1}$, T. G\"uver$^{1}$,
{S. Ak}$^{1}$, 
\newauthor
\"O. \c{C}ak\i rl\i$^{3}$, O. \"Ozdarcan$^{3}$, E. Paunzen$^{4}$, P. De Cat$^{5}$, J. N. Fu$^{6}$, 
\newauthor
Y. Zhang$^{7}$, Y. Hou$^{7}$, G. Li$^{8}$, Y. Wang$^{7}$, W. Zhang$^{8}$,
J. Shi$^{8}$, Y. Wu$^{8}$
\\
  $^1$Istanbul University, Faculty of Science, Department of Astronomy and Space
Sciences, 34119, University-Istanbul, Turkey\\
  $^2$Istanbul University, Graduate School of Science and
Engineering, Department of Astronomy and Space Sciences, 34116\\
Beyaz\i t-Istanbul, Turkey\\
  $^3$Ege University, Science Faculty, Astronomy and Space Sciences
Department, 35100 Bornova, \.Izmir, Turkey\\
  $^4$Department of Theoretical Physics and Astrophysics, Masaryk
University, Kotl\'a\u rsk\'a 2, 611 37 Brno, Czech Republic\\
  $^5$Royal Observatory of Belgium, Ringlaan 3, 1180 Brussel, Belgium\\
  $^6$Department of Astronomy, Beijing Normal University, Beijing 100875, China\\
  $^7$Nanjing Institute of Astronomical Optics \& Technology,
National Astronomical Observatories, Chinese Academy of Sciences, \\
Nanjing 210042, China\\
  $^8$Key Laboratory of Optical Astronomy, National Astronomical
Observatories, Chinese Academy of Sciences,\\
Beijing 100012, China\\
}

\date{}

\pagerange{\pageref{firstpage}--\pageref{lastpage}} \pubyear{}

\begin{document}

\maketitle

\label{firstpage} \begin{abstract}
We present CCD $UBVRI$ photometry of
the field of the open cluster NGC~6866. Structural parameters of the
cluster are determined utilizing the stellar density profile of the
stars in the field. We calculate the probabilities of the stars being a
physical member of the cluster using their astrometric data and perform
further analyses using only the most probable members. The reddening and 
metallicity of the cluster were determined by independent methods. The
LAMOST spectra and the ultraviolet excess of the F and G type
main-sequence stars in the cluster indicate that the metallicity of
the cluster is about the solar value. We estimated the reddening 
$E(B-V)= 0.074 \pm 0.050$ mag using the $U-B$ vs $B-V$ two-colour
diagram. The distance modula, the distance and the age of NGC~6866 were
derived as $\mu = 10.60 \pm 0.10$ mag, $d=1189 \pm 75$ pc and $t = 813
\pm 50$ Myr, respectively, by fitting colour-magnitude diagrams of the
cluster with the PARSEC isochrones. The Galactic orbit of NGC 6866
indicates that the cluster is orbiting in a slightly eccentric
orbit with $e=0.12$. The mass function slope $x=1.35 \pm 0.08$ was
derived by using the most probable members of the cluster. 
\end{abstract}


\section{Introduction}

Distributed along the Galactic disc, open clusters are members of the
thin-disc component of the Milky Way Galaxy. Since the stars in an
open cluster are almost at the same metallicity, distance and age, but
from different spectral classes, these systems provide valuable data
to study the Galactic structure, chemical composition, stellar
structure and star formation processes \citep{Friel1995,Friel2013}. 
Thus, the structural and astrophysical parameters of open clusters 
such as the reddening, metallicity, distance and age have to be 
determined through precise photometric and spectroscopic observations. 
Furthermore, the stellar structure and evolution models exhibit 
considerable discrepancies at different parts of the observational 
main-sequence due to adoption of different input physics \citep{Thompson2014}. 
Since open clusters allow for direct comparisons of the predictions of 
theoretical models with observations, they provide very important tools 
to better understand the physical reasons behind some of these discrepancies.

The most common method used to infer the astrophysical parameters of
the Galactic and extra-galactic open clusters is based on fitting
stellar isochrones to the observed colour-magnitude diagrams (CMDs) and
two-colour diagrams (TCDs) of the clusters. In this method, astrophysical 
parameters are estimated simultaneously via Bayesian statistics. However, 
such simultaneous statistical solutions based on comparison of stellar 
isochrones with the photometric observations suffer from degeneracies 
between parameters, causing large uncertainties in the measured reddening, 
metallicity \citep{Kingetal2005,deMeule2013,Janes2014}, and therefore the 
age values.

Different approaches were suggested to break the age-reddening degeneracy 
in the simultaneous solutions. The main idea behind most of these 
suggestions is to use the maximum available wavelength range, preferably 
including at least one near-infrared (NIR) band 
\citep[cf.][]{Anders2004,Brid2008,Bilir10,deMeule2013}. \cite{Anders2004} 
suggest that, if NIR data cannot be obtained and only observations in
four passbands are available, one of the best photometric band
combination for a reliable parameter determination is $UBVI$. Besides
these suggestions, independent (traditional) and reliable methods that
were developed for the determination of the reddening and metallicity,
such as those recently used by \cite{Yontan2015}, can be used to
constrain these parameters.

Here, we focus on the CCD {\it UBVRI} observations of open cluster
NGC~6866 ($l=79^{\circ}.560, b=+6^{\circ}.839$) since it is one of the
clusters located in the Kepler field \citep{Borucki2011} and
atmospheric model parameters and radial velocities for a considerable
number of stars in the cluster's field are present in the Large Sky
Area Multi-Object Fiber Spectroscopic Telescope
\citep[LAMOST;][]{Luoetal2012} survey. The availability of photometric
and spectroscopic data allow us to calculate the cluster's precise
astrophysical and kinematical parameters.  Following independent
methods we find reddening and metallicity of the cluster from the
stars with high membership probabilities and then keeping these two
constant, we derive its distance modulus and age by fitting stellar
isochrones to the observed CMDs.  Although it has recently been
investigated photometrically, NGC~6866 is a relatively less well-known
cluster \citep{Janes2014} as compared to the other open clusters
located in the Kepler field (NGC~6811, NGC~6819 and NGC~6791). The
colour excesses, distance moduli, distances and ages obtained for
NGC~6866 in the previous studies are summarized in Table 1. Almost all
of these studies have assumed a solar metallicity for the
cluster. \cite{Meretal2008} measured the radial velocities of two
giant stars located in the cluster field and estimated a mean radial
velocity of $V_{r}=+13.68\pm 0.09$ km s$^{-1}$. \cite{FrinMaj2008}
derived a radial velocity of $V_{r}=+12.18$ $\pm$ 0.75 km s$^{-1}$, a
value in agreement with the mean velocity given by
\cite{Meretal2008}. They also found the proper motion components as
$\mu_{\alpha}\cos{\delta}=5.52\pm1.17$ and $\mu_{\delta}=7.97\pm1.09$
mas yr$^{-1}$ for NGC~6866. \cite{Moletal2009} detected 19 variable
stars in the cluster field, changing $V$ apparent magnitudes from
12.1 to 17.4 and $B-V$ colours from 0.3 to 2.2
mag. \cite{Joshietal2012} found 28 periodic variables with $V$
apparent magnitudes between 11.5 and 19.3 and $B-V$ colours between
0.3 and 2.1 mag. Nineteen of these systems were newly identified.

\begin{table}
\setlength{\tabcolsep}{5pt}
\begin{center}
\caption{Colour excesses ($E(B-V)$), distance moduli ($\mu$),
distances ($d$) and ages ($t$) collected from the literature
for the open cluster NGC~6866. References are given in the
last column.}
\begin{tabular}{ccccc}
\hline
$E(B-V)$ &  $\mu$  &    $d$      &    $t$    & Refs.   \\
  (mag)  &  (mag)  &   (pc)      &   (Myr)   &         \\
\hline
0.14     & 10.82   &   1200      &    $-$    &   1, 2   \\
$-$      & $-$     &   $-$       &    230    &   3      \\
0.16     & 11.10   &   $-$       &    $-$    &   4      \\
$-$      & $-$     &   $-$       &    650    &   5      \\
0.17     & 11.33   &   1450      &    480    &   6      \\
0.14     & 11.13   &   1380      &    560    &   7      \\
0.10     & 11.15   &   1470      &    630    &   8      \\
0.19     & 11.08   &   1650      &    800    &   9      \\
0.16     & 10.98   &   1250      &    705    &   10     \\
\hline
\end{tabular}
\\(1) \cite{Hoagetal1961}, (2) \cite{Jonsonetal1961}, (3) \cite{Lind1968},
(4) \cite{HidSut1972}, (5) \cite{LokMat1994}, (6) \cite{Kharetal2005},
(7) \cite{Frolovetal2010}, (8) \cite{Joshietal2012},
(9) \cite{Gunetal2012}, (10) \cite{Janes2014}.
\end{center}
\end{table}

In this work, our main goal is to overcome in part the parameter degeneracy in
the statistical solutions of the astrophysical parameters of
NGC~6866. In Section 2, we summarize the observations and reductions.
We present the CMDs, structural parameters of NGC~6866, and the
membership probabilities of the stars in the cluster field in Section
3. In section 4, we measure the astrophysical parameters of the
cluster. Section 5 discusses the results of the study. 

\section{Observations}

\subsection{Photometry}

CCD $UBVRI$ images of the open cluster NGC~6866 and standard stars
selected from \cite{Land2009} were acquired on 18th July 2012 using a
1m Ritchey-Chr\'etien telescope (T100) located at the T\"UB\.ITAK
National Observatory (TUG)\footnote{www.tug.tubitak.gov.tr} in
Bak{\i}rl{\i}tepe, Antalya/Turkey. With a relatively large CCD
  camera which has a field of view of about 21.5$\times$21.5 arcmin,
  this telescope is very useful for open cluster observations. Since
the observations of NGC~6866 (Fig. 1) and NGC~6811 were performed in
the same night, the details of the observations and photometric
reductions can be found in \cite{Yontan2015}, where the photometric
analyses of NGC~6811 are discussed. Below, a brief summary is given.

\begin{figure}
\begin{center}
\frame{\includegraphics[scale=0.3, angle=0]{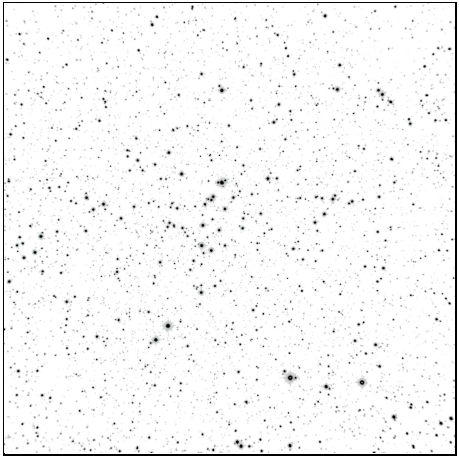}}
\caption[] {\small A 30 second $R$-band image of NGC 6866 obtained
with T100 telescope of the T\"UB\.ITAK National Observatory. The
field of view is about 21$\times$21 arcmin (North top and East left).}
\end{center}
\end{figure}

Short and long exposure images were obtained in each filter of the
cluster's field in order to cover the widest possible flux range. The
night was moderately photometric with a mean seeing of 1$''$.5. A log
of observations is given in Table 2. IRAF\footnote{IRAF is
  distributed by the National Optical Astronomy Observatories.}
routines were used for pre-reduction processes of images and
transforming the pixel coordinates of the objects identified in frames
to equatorial coordinates. The instrumental magnitudes of the standard
stars were measured utilizing the aperture photometry packages of
IRAF. Atmospheric extinction and transformation coefficients for the
observing system were determined from the observations of the standard
stars \citep{Land2009} through the equations given in
\cite{Janes2013}, which are a set of calibration equations to
  transform instrumental magnitudes to the Johnson-Cousins magnitude
  system. For the list of the coefficients for that particular night,
see Table 2 of \cite{Yontan2015}. Source Extractor
(SExtractor)\footnote{SExtractor: Software for source extraction.} and
isophotal photometry \citep{BertArn1996} were performed for the cluster's
field. Transformation of instrumental magnitudes to standard magnitudes 
was done as described in \cite{Yontan2015}.

\begin{table}
\caption{Log of observations, with exposure times
for each passband. $N$ denotes the number of exposure.}
\begin{center}
\begin{tabular}{ccc}
\hline
Filter & Long Exp. time (s)$\times$N & Short Exp. time (s)$\times$N \\
 \hline
$U$ & 180$\times$3 & 90$\times$3 \\
$B$ & ~90$\times$3 & 20$\times$3 \\
$V$ & ~60$\times$3 & ~5$\times$3 \\
$R$ & ~30$\times$3 & ~4$\times$3 \\
$I$ & ~30$\times$3 & ~4$\times$3 \\
\hline
\end{tabular}
\end{center}
\end{table}

\subsection{Spectroscopy}

Spectra of the stars in the direction of NGC 6866 were obtained 
  as part of the LAMOST survey
\citep{Luoetal2012,Cuietal2012,DeCatetal2014,DeCatetal2015a,DeCatetal2015b,
  Renetal2015}. The LAMOST, also called the Guo Jing Telescope, is
located at the Xinglong observatory, China. The telescope has an
aperture of 4m. Its focal plane, which has a 5$^{\circ}$ field of view
in diameter, is covered with 4000 optical fibers connected to 16 sets
of multi-objective optical spectrometers with 250 optical fibers each
\citep{Wangetal1996,Xingetal1998}. The spectrometer obtains low
resolution spectra ($R\simeq 1800$) in two wavelength regions,
3700-5900 and 5700-9000 \AA, using two CCD cameras. LAMOST can take
spectra of all objects in the field of view down to a magnitude of
$V\sim$ 17.8 simultaneously. Its capability of tracking the
motion of celestial objects during about four hours while they are
passing the meridian makes the LAMOST an ideal instrument to collect
low-resolution spectra of the objects in the Kepler field. 
  Spectra obtained with the LAMOST are wavelength calibrated and
  intensity normalized following the procedures detailed in
  \cite{Luoetal2012}. For detailed information about the LAMOST, see
\cite{Cuietal2012} and \cite{Zhaoetal2012}.

 We scanned the LAMOST database in order to find the stars with
  spectra, which are located in our field of view, and found 31 stars
  for which atmospheric parameters and radial velocities were
  measured. We presented the KIC names and equatorial coordinates of
  these stars in Table 3 (see also, Fig. 7).

The stellar atmospheric parameters, e.g. effective temperature
  $T_{eff}$, surface gravity $\log g$, and metallicity $[Fe/H]$, are
  estimated via the stellar parameters pipeline of the LAMOST. The
  spectral data reduction procedure consists of three stages: $i)$
  LAMOST 2D pipeline works on reduction of CCD data including dark and
  bias subtraction, flat field correction, spectra extraction, sky
  subtraction, wavelength calibration, sub-exposure merging and
  wavelength band combination. $ii)$ LAMOST 1D pipeline identifies
  spectral type classes and measures radial velocities for star or the
  redshifts for galaxy or QSO using cross correlation. The pipeline
  produces four prime classifications, namely STAR (with released
  subclass), GALAXY, QSO and UNKNOWN, $iii)$ LAMOST Stellar Parameter
  pipeline \citep[LASP;][]{Wuetal2015, Luoetal15}, automatically
  derives the stellar parameters for late A and FGK type stellar
  observation from LAMOST survey stars. The LASP consecutively adopts
  correlation function interpolation \citep[CFI;][]{Duetal12} method
  and Universit\'e de Lyon Spectroscopic analysis Software
  \citep[ULySS;][]{Koletal2009,Wu09, Wuetal2011} method to determine
  the stellar parameters. The LASP first uses CFI method to get an
  initial coarse guess for $T_{eff}$, $\log g$, $[Fe/H]$ that serve as
  the starting values for ULySS, then ULySS determines the
  parameters by minimizing the $\chi^2$ values between the observation
  and the template spectra which were generated by an interpolator
  built on the empirical ELODIE stellar library \citep{Prugniel01}.
  Details for the spectral analysis and propagated errors can be found
  in \citet{Luoetal15}.

\begin{table*}
\setlength{\tabcolsep}{5pt}
\begin{center}
\caption{ID numbers, Kepler Input Catalogue (KIC) names and equatorial 
coordinates of the stars with LAMOST spectra in the direction of 
NGC 6866. Apparent magnitudes ($V$), colour indices ($B-V$), observing 
dates and signal-to-noise ratio (SNR) of the spectra are also given.}
\begin{tabular}{cccccccc}
\hline
ID & KIC & $\alpha_{2000.0}$ &    $\delta_{2000.0}$    & $V$    & $B-V$  &  Date& SNR  \\
   &     &    (hh:mm:ss.ss)  & ($^{\circ}$~~$'$~~$''$) &  (mag) & (mag)  &      &      \\
\hline
 1    & 8329578 & 20:03:00.39 & +44:16:54.13 & 13.519 & 0.728 & 25.10.2013 & 78 \\
 2    & 8329629 & 20:03:04.09 & +44:14:15.65 & 12.588 & 0.267 & 25.09.2013 & 186 \\
 3    & 8128952 & 20:03:04.83 & +43:59:58.60 & 13.786 & 0.568 & 25.09.2013 & 74 \\
 4    & 8263825 & 20:03:12.47 & +44:06:59.03 & 13.445 & 0.842 & 25.10.2013 & 87 \\
 5    & 8329888 & 20:03:22.22 & +44:16:57.95 & 12.162 & 0.295 & 25.10.2013 & 182 \\
 6    & 8329894 & 20:03:22.81 & +44:15:50.46 & 11.578 & 1.030 & 25.09.2013 & 166 \\
 7    & 8264037 & 20:03:26.11 & +44:10:05.45 & 11.755 & 0.356 & 25.09.2013 & 208 \\
 8    & 8264075 & 20:03:28.33 & +44:07:55.26 & 13.784 & 0.393 & 25.10.2013 & 66 \\
 9    & 8264148 & 20:03:32.47 & +44:06:46.17 & 14.032 & 0.479 & 25.09.2013 & 60 \\
10    & 8330092 & 20:03:34.92 & +44:14:50.16 & 13.548 & 0.439 & 25.09.2013 & 89 \\
11    & 8197368 & 20:03:44.34 & +44:05:39.30 & 12.777 & 0.475 & 25.10.2013 & 137 \\
12    & 8330251 & 20:03:45.04 & +44:15:19.86 & 13.321 & 0.688 & 25.09.2013 & 102 \\
13    & 8129588 & 20:03:45.24 & +43:59:22.51 & 13.207 & 0.341 & 25.10.2013 & 114 \\
14    & 8197440 & 20:03:48.06 & +44:02:48.38 & 12.943 & 0.674 & 25.10.2013 & 137 \\
15    & 8264534 & 20:03:54.16 & +44:06:46.04 & 12.628 & 0.222 & 25.09.2013 & 169 \\
16    & 8264581 & 20:03:57.12 & +44:08:16.86 & 13.509 & 0.436 & 25.09.2013 & 103 \\
17    & 8330453 & 20:03:57.80 & +44:16:32.51 & 12.804 & 0.738 & 25.09.2013 & 131 \\
18    & 8264617 & 20:03:59.33 & +44:10:25.84 & 13.975 & 0.439 & 25.09.2013 & 79 \\
19    & 8264674 & 20:04:02.84 & +44:11:55.63 & 11.164 & 0.251 & 25.09.2013 & 328 \\
20    & 8330543 & 20:04:03.64 & +44:15:34.07 & 13.219 & 0.551 & 25.10.2013 & 99 \\
21    & 8264698 & 20:04:03.95 & +44:10:20.72 & 12.356 & 0.318 & 25.09.2013 & 197 \\
22    & 8197761 & 20:04:09.30 & +44:04:16.39 & 10.639 & 0.338 & 25.09.2013 & 329 \\
23    & 8396247 & 20:04:09.36 & +44:19:10.03 & 13.835 & 1.852 & 25.09.2013 & 36 \\
24    & 8330778 & 20:04:16.17 & +44:12:04.66 & 13.468 & 0.379 & 25.10.2013 & 118 \\
25    & 8330790 & 20:04:16.73 & +44:12:43.66 & 13.143 & 0.314 & 25.09.2013 & 161 \\
26    & 8264949 & 20:04:18.36 & +44:09:52.47 & 12.122 & 0.302 & 25.09.2013 & 235 \\
27    & 8265068 & 20:04:25.50 & +44:10:16.54 & 12.029 & 0.333 & 25.09.2013 & 282 \\
28    & 8198114 & 20:04:32.20 & +44:05:12.84 & 13.728 & 0.660 & 25.09.2013 & 71 \\
20    & 8265356 & 20:04:43.26 & +44:08:02.05 & 12.367 & 0.611 & 17.10.2013 & 93 \\
30    & 8331290 & 20:04:44.09 & +44:15:05.77 & 14.059 & 0.691 & 25.09.2013 & 54 \\
31    & 8265377 & 20:04:44.37 & +44:08:58.53 & 13.170 & 0.315 & 25.09.2013 & 117 \\
\hline
\end{tabular}
\end{center}
\end{table*}

\begin{table*}
\setlength{\tabcolsep}{2pt}
\begin{center}
\small{
\caption{Photometric and astrometric catalogue for the open cluster NGC 6866.
The complete table can be obtained electronically.}
\begin{tabular}{ccccccccccc}
\hline
        ID &         $\alpha_{2000}$ &   $\delta_{2000}$ &          $V$ &        $U-B$ &        $B-V$ &        $V-R$ &        $R-I$ & $\mu_{\alpha}\cos \delta$&   $\mu_{\delta}$ &  $P$ \\
          &(hh:mm:ss.ss)&(dd:mm:ss.ss)&          (mag) &        (mag) &        (mag) &        (mag) &        (mag) & (mas yr$^{-1}$)&   (mas yr$^{-1}$) &   (\%)      \\
\hline
    1     & 20:02:57.52 & +44:11:17.39 & 18.454$\pm$0.017 & 1.445$\pm$0.270 & 1.381$\pm$0.036 & 0.902$\pm$0.022 & 0.676$\pm$0.019 & $--$ & $--$ & $--$ \\
    2     & 20:02:57.53 & +44:04:17.23 & 18.546$\pm$0.018 & 1.163$\pm$0.182 & 1.149$\pm$0.033 & 0.685$\pm$0.024 & 0.697$\pm$0.023 & $--$ & $--$ & $--$ \\
    3     & 20:02:57.58 & +44:14:51.51 & 18.319$\pm$0.015 & 0.703$\pm$0.097 & 1.033$\pm$0.026 & 0.637$\pm$0.021 & 0.643$\pm$0.021 & $--$ & $--$ & $--$ \\
    4     & 20:02:57.74 & +44:09:23.85 & 18.873$\pm$0.021 & 0.849$\pm$0.154 & 1.082$\pm$0.037 & 0.623$\pm$0.029 & 0.743$\pm$0.028 & $--$ & $--$ & $--$ \\
    5     & 20:02:57.77 & +44:12:06.25 & 16.612$\pm$0.005 & 0.607$\pm$0.029 & 1.002$\pm$0.008 & 0.607$\pm$0.007 & 0.611$\pm$0.007 & 9.0$\pm$6.7 & 1.2$\pm$6.7 & 32 \\
       .. & ........... & ...........  & ................ & ............... & ............... & ............... & ............... & ............ & ............. &       .... \\
\hline
\end{tabular}
}
\end{center}
\end{table*}

\section{Data analysis}

\subsection{Identification of stars and photometric errors}

We identified 2096 sources in the field of NGC~6866 and
constructed a photometric and astrometric catalogue. The stellarity 
index (SI) provided by SExtractor \citep{BertArn1996} was used to detect 
non-stellar objects, most likely galaxies, in our catalogue. The SI 
has values between 0 and 1. According to \cite{BertArn1996}, a source with
the SI close to 1 is a point source (most probably a star), while an 
extended object has an SI close to zero. Application of the SI to the 
stellar fields can be found in \cite{Andetal2002} and \cite{Karetal2004} 
who showed that the objects with an SI smaller than 0.8 can be assumed to 
be extended objects. Thus we adopted that the objects with an SI larger 
than 0.8 are most probably stars. The $V$ apparent magnitude versus SI 
diagram of 2096 objects is plotted in Fig. 2. Resulting catalogue contains 
2089 stars. Individual stars in the final photometric catalogue are 
tabulated in Table 4. The columns of the table are organized as ID, 
equatorial coordinates, $V$ apparent magnitude, colours, proper motion 
components and the probability of membership. The proper 
motions of the stars were taken from the astrometric catalogue of 
\citet[UCAC4;][]{Zachetal13}.

\begin{figure}
\begin{center}
\includegraphics[scale=0.45, angle=0]{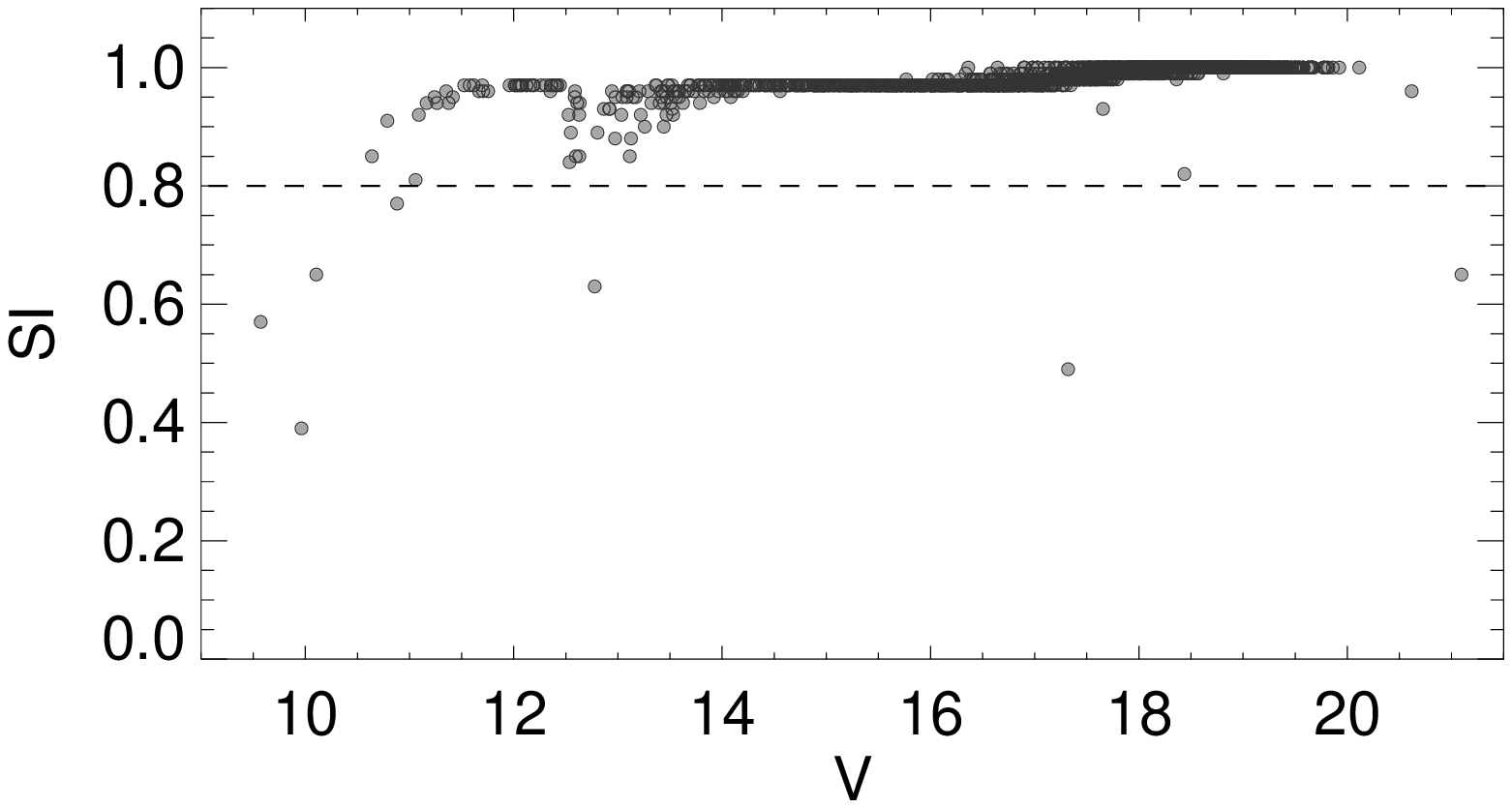}
\caption[] {\small Stellarity index (SI) for 2096 sources observed
in the direction to the cluster NGC 6866, as a function of $V$
apparent magnitude. Dashed line represents the SI of 0.8. Only
seven sources with SI $<$ 0.8 are classified as non-stellar sources.}
\end{center}
\end{figure}
\begin{table}
\setlength{\tabcolsep}{3.5pt}
\caption{Mean errors of the photometric measurements for the stars in the direction
of NGC 6866. $N$ indicates the number of stars within the $V$ apparent magnitude
range given in the first column.}
\begin{center}
\begin{tabular}{ccccccc}
\hline
Mag. Range &    $N$ &  $\sigma_V$ & $\sigma_{U-B}$ &  $\sigma_{B-V}$
& $\sigma_{V-R}$ &  $\sigma_{R-I}$\\
\hline
$10<V\leq12$ & 18  & 0.001 & 0.002 & 0.002 & 0.002 & 0.002 \\
$12<V\leq13$ & 39  & 0.001 & 0.002 & 0.002 & 0.002 & 0.002 \\
$13<V\leq14$ & 65  & 0.001 & 0.004 & 0.001 & 0.002 & 0.002 \\
$14<V\leq15$ & 109 & 0.002 & 0.011 & 0.003 & 0.002 & 0.003 \\
$15<V\leq16$ & 192 & 0.003 & 0.022 & 0.005 & 0.004 & 0.004 \\
$16<V\leq17$ & 320 & 0.005 & 0.039 & 0.009 & 0.007 & 0.007 \\
$17<V\leq18$ & 560 & 0.009 & 0.083 & 0.016 & 0.013 & 0.012 \\
$18<V\leq19$ & 608 & 0.016 & 0.150 & 0.030 & 0.023 & 0.021 \\
$19<V\leq20$ & 178 & 0.027 & 0.183 & 0.050 & 0.037 & 0.035 \\
\hline
\end{tabular}
\end{center}
\end{table}

The errors of the measurements in the $V$ band and $U-B$,
$B-V$, $V-R$, $V-I$ and $R-I$ colours are shown in Fig. 3
as a function of the apparent $V$ magnitude. Mean errors in the selected 
magnitude ranges are listed in Table 5. The errors are relatively small 
for stars with $V<18~\rm{mag}$, while they increase exponentially towards 
fainter magnitudes. Thus, we decided to use stars with $V<18$ mag for 
further analysis. With this selection, the number of remaining stars for 
analysis is 1301 in the field of NGC~6866. As expected, the largest errors 
for a given $V$ magnitude occurred in the $U-B$ colours of the stars. 

\begin{figure}
\begin{center}
\includegraphics[trim=0.2cm 0.5cm 0cm 0cm, clip=true, scale=0.8]{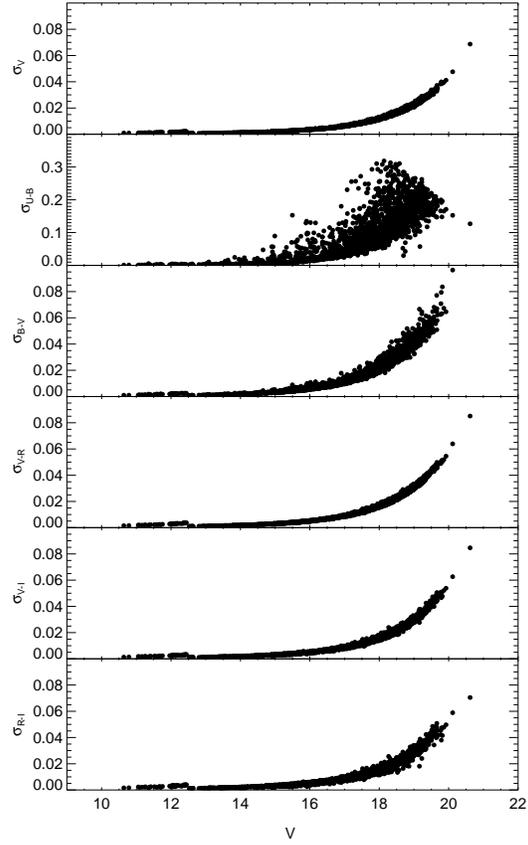}
\caption[] {\small $V$ apparent magnitude versus colour and magnitude errors
for the stars identified in the field of the open cluster NGC~6866.}
\end{center}
\end{figure}

\begin{figure}
\begin{center}
\includegraphics[scale=0.5, angle=0]{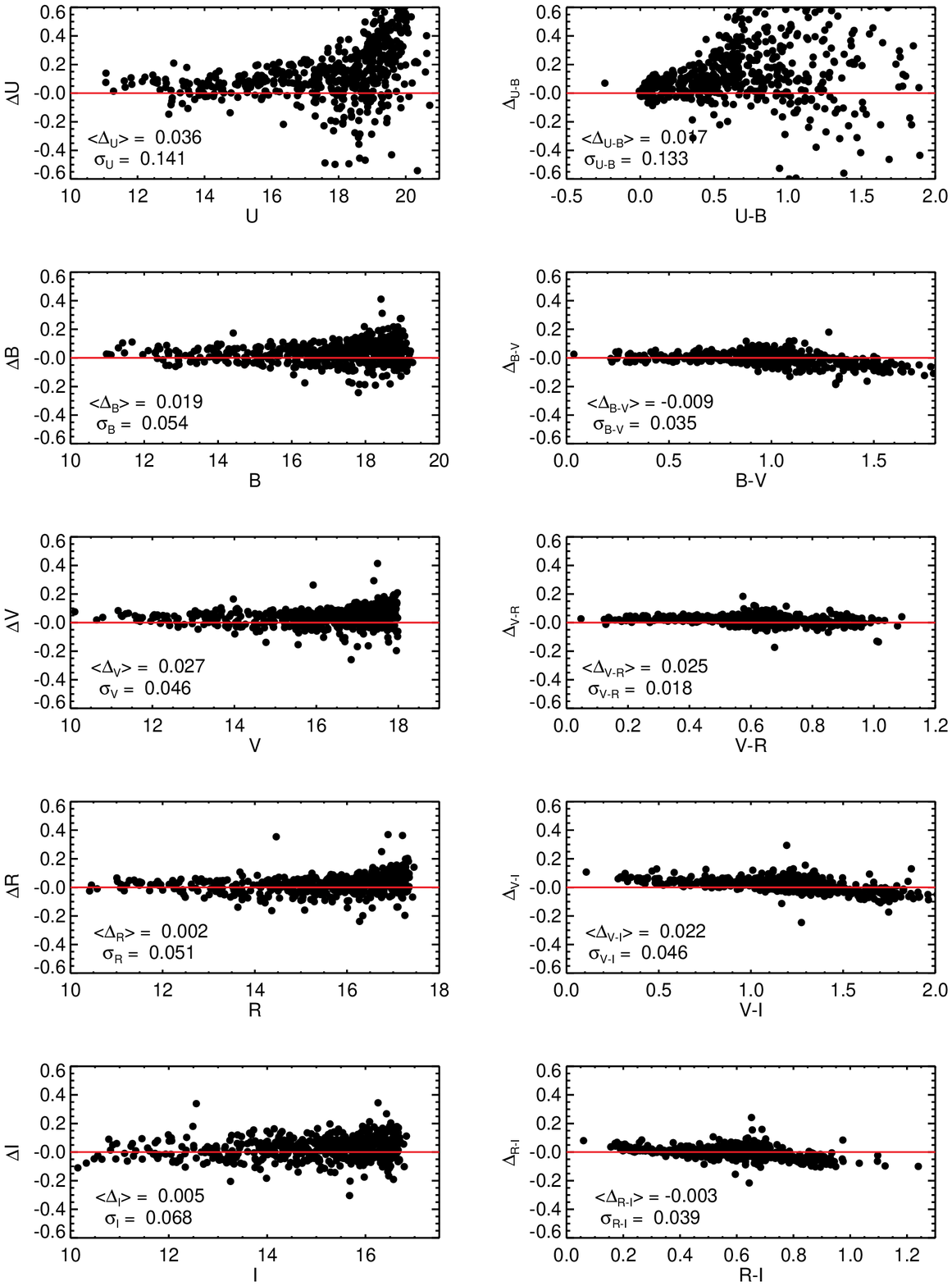}
\caption[] {\small  Comparison of the magnitudes and colours for the stars
observed both in this study and in \cite{Joshietal2012}. The means and standard
deviations of the differences are shown in panels.}
\end{center}
\end{figure}

  We compared our photometric measurements with those of
  \cite{Joshietal2012} and \cite{Janes2014} who utilized the same
  photometric bands, i.e. $UBVRI$, in their investigations.  A cross-match of
  our catalogue with these two catalogues resulted in 800 stars for
  \cite{Joshietal2012} and 429 stars for \cite{Janes2014}. Comparisons
  between our and their datasets are shown in Fig. 4 and Fig. 5.  In
  these figures, values on the abscissae refer to our measurements,
  while the magnitude or colour differences in the ordinates present
  the differences between the our and their catalogues. The means and
  the standard deviations of the magnitudes and colours obtained from
  the differences are small except for the standard deviations of
  $\Delta_{U}$ and $\Delta_{U-B}$, i. e. $\sigma_{U}= 0.141$ and
  $\sigma_{U-B}= 0.133$ mag in the comparison with the
  \cite{Joshietal2012} and $\sigma_{U}= 0.108$ and $\sigma_{U-B}=
  0.095$ mag in the comparison with the \cite{Janes2014}. Note
  that the large mean differences and standard deviations calculated
  for the $U$ magnitudes and $U-B$ colours are due to faint field
  stars beyond $U\sim 17$ mag, for which the data are not used in the
  derivation of the cluster parameters. Agreement between our study
  and \cite{Janes2014} is much better than that with the study of
  \cite{Joshietal2012}.

\subsection{Cluster radius and radial stellar surface density}

We estimated the stellar density profile of the open cluster NGC~6866
using stars with $V<18$ mag in the field. The central coordinates of
the cluster were assumed to be as given in
WEBDA\footnote{http://www.univie.ac.at/webda/} database
($\alpha_{2000.0}=20^{h}03^{m}55^{s}$,
$\delta_{2000.0}=+44^{\circ}09'30''$). We then calculated the stellar
density in an area defined by a circle with a radius of 1 arcmin and
centered on these coordinates. From this central circle, we calculated
the variation of stellar density using annuli with width of 2
arcmin. The last annulus had a width of 3 arcmin because of a
significant decrease in the number of stars. The resulting stellar
density profile of the cluster is plotted in Fig. 6.

We fitted this density profile with the \cite{King1962} model
defined as:

\begin{eqnarray}
  \rho(r)=f_{bg}+\frac{f_0}{1+(r/r_{c})^2},
\end{eqnarray}
where $r$ is the radius of the cluster centered at the celestial
coordinates given above. $f_{bg}$, $f_0$ and $r_c$ denote the
background stellar density, the central stellar density and the
core radius of the cluster, respectively. We fitted the King model
to the observed radial density profile for NGC~6866 and used
a $\chi^{2}$ minimization technique to determine $f_{bg}$, $f_0$
and $r_c$. The best fit to the density profile is shown with
a solid line in Fig. 6. We estimated the central stellar density 
and core radius of the cluster, and the background stellar density as
$f_{0}=2.28$ $\pm$ 0.02 stars arcmin$^{-2}$, $r_{c}=3.24$ $\pm$ 0.04 
arcmin and $f_{bg}=5.33$ $\pm$ 0.01 stars arcmin$^{-2}$, respectively.

\cite{Joshietal2012} estimated the central stellar density and core
radius of the cluster, and the background stellar density as
$f_{0}=5.7$ $\pm$ 0.7 stars arcmin$^{-2}$, $r_{c}=2.0$ $\pm$ 0.5
arcmin and $f_{bg}=7.85$ stars arcmin$^{-2}$, respectively, while
\cite{Janes2014} found $f_{0}=1.65$ stars arcmin$^{-2}$, $r_{c}=4.70$
arcmin and $f_{bg}=0.63$ stars arcmin$^{-2}$, respectively.
Structural parameters in this study are generally in agreement with
those in \cite{Joshietal2012} whereas they are not in agreement with
those in \cite{Janes2014}. This is most likely caused by the selection
made by \cite{Janes2014} since they considered only 75 stars of 11-16
mag within a circular field of 3 arcmin radius from the cluster's
centre.

\begin{figure}
\begin{center}
\includegraphics[scale=0.5, angle=0]{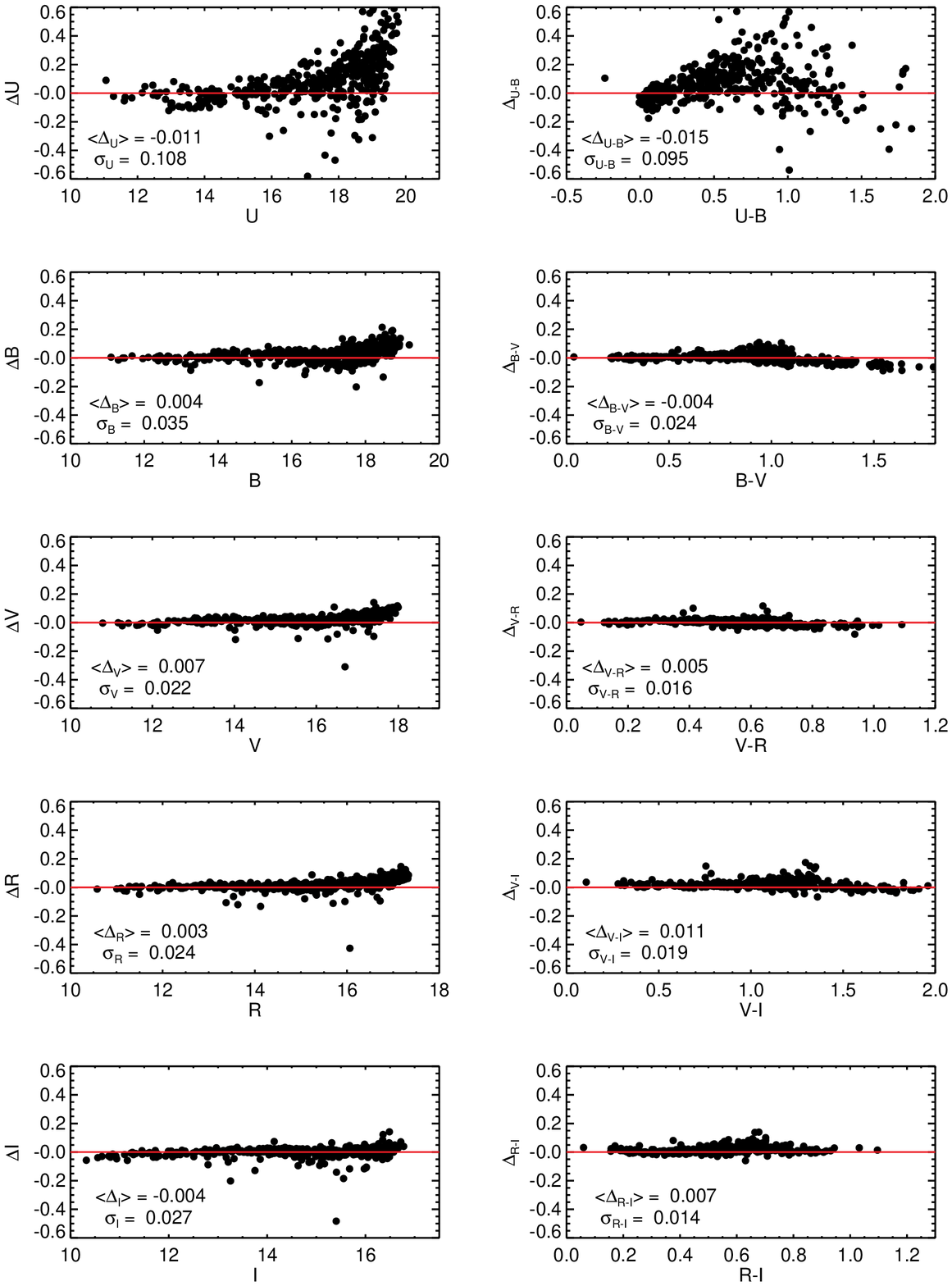}
\caption[] {Same as Fig.4 but for \cite{Janes2014}.}
\end{center}
\end{figure}

\subsection{CMDs and membership probabilities}

In order to derive the parameters of NGC~6866, we used its CMDs in Fig.
7, i.e. $V$ vs $U-B$, $V$ vs $B-V$, $V$ vs $V-R$ and $V$ vs $R-I$.
An inspection by eye suggests that the cluster is rather sparse. Since 
most of the stars brighter than $V=17$ mag are lying along a sequence, 
we conclude that these stars are probably representing the main-sequence 
of the cluster. On the contrary, most of the stars fainter than 
$V=17$ mag are field stars. In order to confirm this conclusion we 
selected a circular area of the core radius and an annulus between 10.5 
and 11 arcmin from the centre of the cluster. We found 237 and 67 stars 
in the circular area and the annulus, respectively, and presented their 
positions in the $V$ vs $B-V$ CMD in Fig. 8 demonstrating that 
most of the stars close to the cluster centre are generally brighter than 
$V\approx$\,17 mag and lying along the cluster's main-sequence, as 
indicated above. However, the stars in the annulus are mostly 
fainter than $V\approx$\,17 mag and they are probably field stars.

\begin{figure}
\begin{center}
\includegraphics[scale=0.45, angle=0]{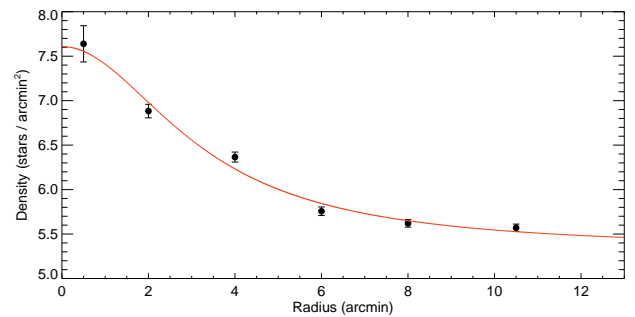}
\caption[] {\small Stellar density profile of NGC 6866.
Errors were determined from sampling statistics:
$1/\sqrt{N}$, where $N$ is the number of stars used in the density
estimation.}
\end{center}
\end{figure}

\begin{figure*}
\includegraphics[trim=0.2cm 0cm 0cm 0cm, clip=true, scale=0.52]{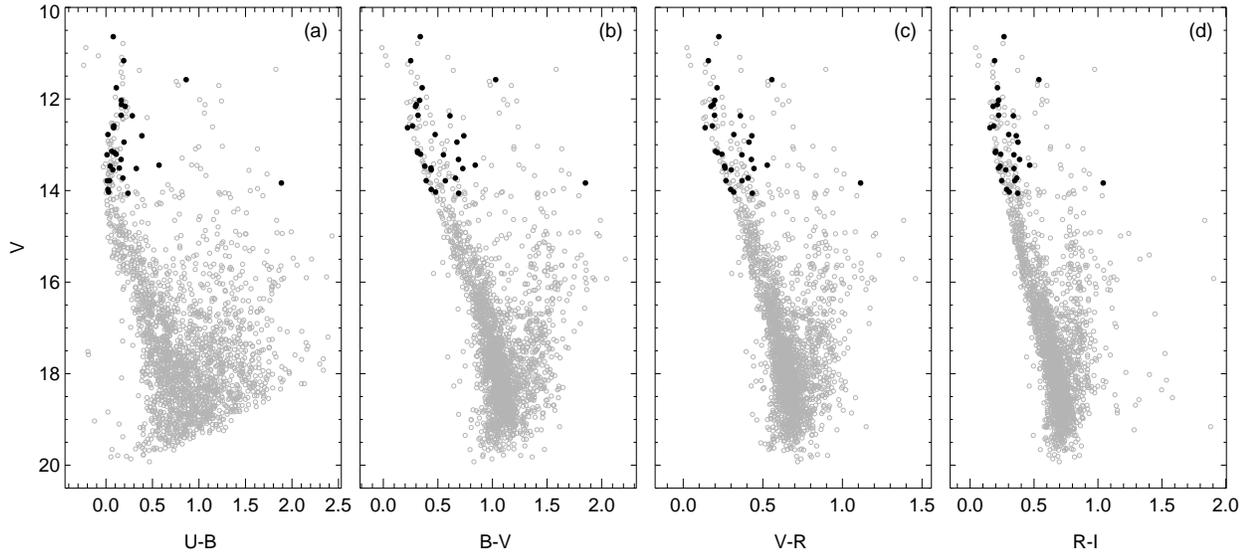}
\caption[] {The CMDs for the cluster NGC 6866. Black dots represent 
the stars with LAMOST spectra.}
\end{figure*}

 For relatively old clusters like NGC~6866, the red clump (RC)
  stars in their CMDs can be very useful to determine their distances
  and ages since they have been utilized as standard candles for
  distance estimates
  \citep[i.e.][]{PacSta1998,Cabetal2005,Cabetal2007,Biletal2013a,
  Karaali2013, Yazetal2013}.
  The RC stars in $V$ vs $B-V$ CMD can be identified in the colour
  range $0.7 \leq (B-V)_{0} \leq 1.2$ mag and the absolute magnitude
  range $0 \leq M_{V} \leq 2$ mag \citep{Biletal2013b}. Here,
$(B-V)_{0}$ denotes the de-reddened $B-V$ colour. Indeed, there are 
stars in these colour and apparent magnitude ranges in Fig. 7. If they 
are members of the cluster, these RC stars can be used to confirm 
the distance estimation of NGC~6866. On the other hand, the position 
of another small group of bright and blue stars in the CMDs suggests 
that the turn-off point of the cluster lie within $11<V<12$ mag and 
 $0.2\leq B-V\leq 0.4$ mag.

\begin{figure}
\begin{center}
\includegraphics[scale=0.5, angle=0]{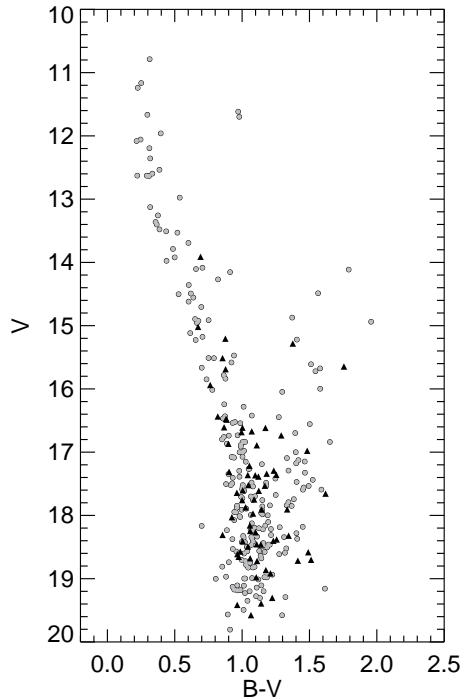}
\caption[] {\small The location of stars in the $V$ vs $B-V$ CMD
    selected from the two different regions in the field of NGC 6866. 
    Gray dots denote the stars located in the region with $r_{c} \leq 
    3.2$ arcmin, while black triangles indicate the stars within an annulus 
    between 10.5 and 11 arcmin from the centre of the cluster.}
\end{center}
\end{figure}

Although identifying some stars near the RC region and turn-off point
of the CMDs is very promising for the analysis, we can not confidently
use these stars for further analysis without knowing if they are
physical members of the cluster. In addition, determining the likely
members of the cluster can be very useful to clearly demonstrate the
cluster's main-sequence. Thus, the probabilities of the stars in the
field being physical members of the cluster must be estimated.  For
the membership probabilities of the individual stars, we employed the
method given by \citet{Bala98}. This method takes both the errors of
the mean cluster and the stellar proper motions into account. The
basic idea of the non-parametric method for the cluster-field
separation in the 2-dimensional proper motion space is the empirical
determination of the cluster and field stars' distributions without
any assumption about their shape. We used the kernel estimation
technique (with a circular Gaussian kernel function) to derive the
data distributions. The proper motions of the stars were taken from
the UCAC4 catalogue.  Additionally, we compared the results with those
of the algorithm published by \citet{Java06}, yielding excellent
agreement. For this, we considered rectangular coordinates of the
stars in the field, measured in two epochs, first of our observations
and second the UCAC4 ones. The histogram of the differences
efficiently discriminate the members from non-members.

In order to determine the most likely members of the cluster we
applied the following procedure. First, we selected only the stars in
a circle with a radius of 6 arcmin whose centre coincides with the
cluster's centre. The selected radius of 6 arcmin corresponds to
  about two times the core radius of the cluster. As can be seen from
  Fig. 6, the field stars are dominant beyond this radius. The median
  value of the membership probabilities of these stars is found as
  $P=50\%$. The histogram of the probabilities is shown in 
  Fig. 9. Second, in order to identify the main-sequence stars
  of the cluster, we fitted the zero age main-sequence (ZAMS) of
  \cite{Sungetal2013} for solar metallicity to the $V$ vs $B-V$ CMD of
  NGC 6866 for 12.75 $\leq V \leq$ 17 mag using only the stars with
  $P>50\%$. The faint apparent magnitude limit was assumed to be
  $V=17$ mag as the proper motions could be obtained only for the stars
  brighter than this magnitude. By shifting the fitted main-sequence to
  brighter $V$ magnitudes by 0.75 mag, a band like region in $V$ vs $B-V$ 
  CMD (see Fig. 10) was obtained to cover the binary stars, as well. 
  However, we found visually out that the stars brighter than $V=12.75$ mag 
  have already left the ZAMS. Thus, we conclude that this magnitude roughly
  corresponds to the turn-off point of the cluster. Hence, we assumed
  that all stars brighter than $V=$ 12.75 mag, which are located in
  the circle defined above and having a probability of membership
  larger than $P=50\%$, are the most likely members of the
  cluster. Finally, we assumed that all stars with a membership
  probability $P>50\%$ and located within the band-like region defined
  above are the most likely members of NGC~6866 on the main-sequence.
  With this procedure, we identified 64 stars which are used for
  further analyses. These stars are indicated with black dots in
  Fig. 10.

\begin{figure}
\begin{center}
\includegraphics[trim=0cm 9cm 0cm 9cm, clip=true, scale=0.40]{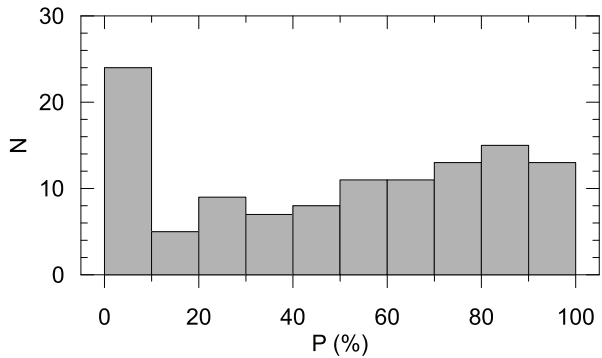}
\caption[] {\small The histogram of the membership probabilities
estimated for the stars which are located in a circle with a radius
of 6 arcmin from the centre of NGC 6866.}
\end{center}
\end{figure}

\begin{figure}
\begin{center}
\includegraphics[scale=0.50, angle=0]{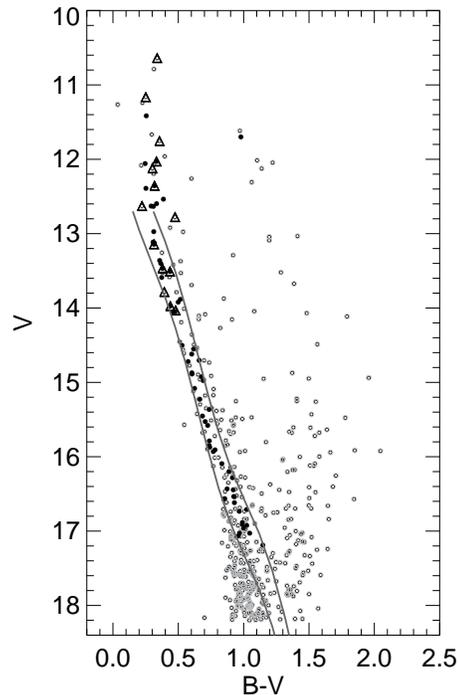}
\caption[] {\small $V$ vs $B-V$ CMD of NGC 6866 constructed
using stars which are located in a circle of 6 arcmin
radius from the centre of the cluster. Solid lines represent the ZAMS
of \cite{Sungetal2013} and the one shifted by an amount of 0.75 mag to
the bright $V$ magnitudes. Black dots denote the most
probable cluster stars that are identified using a procedure
explained in the text. Open triangles indicate the stars with LAMOST
spectra.}
\end{center}
\end{figure}

\section{Determination of the astrophysical parameters of NGC~6866}

\subsection{The reddening}

Before the determination of the metallicity of the cluster using
photometric observations, its reddening should be estimated. Thus, for
the determination of the colour excesses $E(U-B)$ and $E(B-V)$, we
used the 64 probable members of the cluster selected according to the
procedure in Section 3.3. We compared the positions of these stars in
the $U-B$ vs $B-V$ TCD with the ZAMS of \cite{Sungetal2013} with a
solar metallicity. In order to do this, we shifted the
  de-reddened main-sequence curves of \cite{Sungetal2013} within the
  range $0 \leq E(B-V) \leq 0.20$ mag with steps of 0.001~mag until it
  fits well with the $U-B$ vs $B-V$ TCD of NGC 6866. The shift in
the $U-B$ axis was calculated by adopting the following equation
\citep{Hiltner1956}:

\begin{eqnarray}
E(U-B) = E(B-V) \times [0.72 + 0.05 \times E(B-V)].
\end{eqnarray}

To define the goodness of the fit we adopted the minimum $\chi^2$ method. 
Fig. 11 shows the $U-B$ vs $B-V$ TCD of NGC 6866 for the 64
probable members of the cluster. In Fig. 11, the dashed red line
represents the reddened ZAMS of \cite{Sungetal2013} while the black 
dotted line shows the de-reddened ZAMS of the cluster and green lines
the $\pm 1\sigma$ deviations. This method gives the following colour 
excesses: $E(U-B)=0.054 \pm 0.036$ and $E(B-V)=0.074 \pm 0.050$ mag. 
The errors indicate the $\pm 1\sigma$ deviations. In order to evaluate 
the $E(V-R)$ and $E(R-I)$ colour excesses we usedthe following equations 
of \cite{Cardetal1989}:

\begin{eqnarray}
E(V-R) = 0.65 \times E(B-V),\\
E(R-I) = 0.60 \times E(B-V). \nonumber
\end{eqnarray}
The colour excesses calculated are $E(V-R)=0.048 \pm 0.050$ and
$E(R-I)=0.044 \pm 0.050$ mag.

\begin{figure}
\begin{center}
\includegraphics[scale=0.70, angle=0]{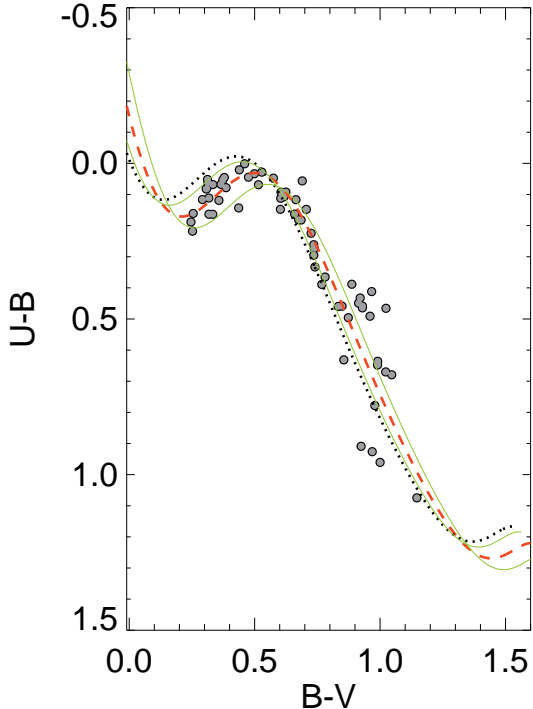}
\caption[] {\small $U-B$ vs $B-V$ TCD for the main-sequence stars with
$12\leq V\leq 17$ mag of NGC 6866. The reddened and de-reddened main-sequence
curves \citep{Sungetal2013} fitted to the cluster stars are shown with red
dashed and black dotted lines, respectively. Green lines represent $\pm 1\sigma$ 
deviations.}
\end{center}
\end{figure}

\subsection{Metallicity and radial velocity from the LAMOST spectra}

Stars with the LAMOST spectra which are located in a circle of 6 arcmin 
radius from the centre of the cluster were used to derive the mean radial 
velocity and the metallicity of the cluster. The effective temperatures 
$T_{eff}$, surface gravities $\log g$, metallicities $[Fe/H]$ and radial 
velocities $V_r$ of these stars are listed in Table 6. These stars are also 
indicated with green circle in Fig. 10.

Median metallicities and radial velocities of 14 stars in Table 6 are
$-0.09\pm 0.11$ dex and $9.86\pm 26.01$ km s$^{-1}$, respectively.
Six stars out of 14 have the membership probabilities larger than
50$\%$. Considering these six stars, we conclude that the median
metallicity and the radial velocity of NGC~6866 are
$[Fe/H]=-0.10\pm 0.13$ dex and $V_r=10.58\pm 31.83$ km s$^{-1}$,
respectively.  Radial velocity of the cluster measured in this study
is in agreement with the radial velocities derived by
\cite{Meretal2008} and \cite{FrinMaj2008}, where using only two stars
in each study, $V_{r}=+13.68\pm 0.09$ and $V_{r}=+12.18\pm 0.75$ km
s$^{-1}$ are given, respectively. We note that a comparison revealed
that there are no common stars between our sample and the stars used
in the other two studies.

\begin{table}
\setlength{\tabcolsep}{4pt}
\begin{center}
\small{
\caption{Results of the spectroscopic analysis of the stars in the direction
to NGC~6866. Membership probabilities ($P$), effective
temperatures ($T_{eff}$), surface gravities ($\log g$), metallicities
($[Fe/H]$) and radial velocities ($V_r$) are given.}
\begin{tabular}{lccccc}
\hline
KIC   &  $P$  &  $T_{eff}$ & $\log g$ & $[Fe/H]$ &  $V_r$  \\
      & ($\%$)  &  (K) & (cgs)  & (dex)   & (km s$^{-1}$) \\
\hline
    8197761 & 00  & 7153$\pm$77  & 4.02$\pm$0.32 & -0.04$\pm$0.09 & -29.12$\pm$24.47 \\
    8264534 & 27  & 8270$\pm$83  & 3.87$\pm$0.34 & -0.03$\pm$0.10 &  -6.48$\pm$20.51 \\
    8264674 & 36  & 8232$\pm$59  & 3.86$\pm$0.32 & -0.12$\pm$0.08 &   9.47$\pm$20.87 \\
    8264698 & 61  & 7730$\pm$64  & 3.98$\pm$0.35 & -0.15$\pm$0.13 &  12.92$\pm$32.76 \\
    8264949 & 00  & 7726$\pm$53  & 3.89$\pm$0.34 & -0.16$\pm$0.11 &  11.59$\pm$29.85 \\
    8264148 & 40  & 6977$\pm$152 & 4.09$\pm$0.33 & -0.06$\pm$0.16 &  14.10$\pm$22.80 \\
    8264581 & 90  & 7337$\pm$113 & 3.66$\pm$0.38 &  0.62$\pm$0.10 &  10.25$\pm$17.00 \\
    8264037 & 00  & 7244$\pm$92  & 3.97$\pm$0.36 & -0.04$\pm$0.11 &   5.29$\pm$29.05 \\
    8330790 & 94  & 7669$\pm$69  & 4.05$\pm$0.33 & -0.09$\pm$0.13 & -22.79$\pm$32.92 \\
    8264617 & 76  & 7177$\pm$139 & 4.11$\pm$0.32 & -0.20$\pm$0.16 &  10.90$\pm$25.28 \\
    8265068 & 75  & 7570$\pm$50  & 3.98$\pm$0.31 & -0.11$\pm$0.11 &  13.59$\pm$30.90 \\
    8264075 & 17  & 7177$\pm$152 & 4.11$\pm$0.34 & -0.12$\pm$0.17 &  10.60$\pm$26.74 \\
    8197368 & 20  & 6622$\pm$107 & 4.24$\pm$0.35 & -0.08$\pm$0.11 &   4.79$\pm$22.63 \\
    8330778 & 53  & 7268$\pm$130 & 3.99$\pm$0.47 & -0.09$\pm$0.19 &   8.45$\pm$59.52 \\
\hline
\end{tabular}
}
\end{center}
\end{table}

\subsection{Photometric metallicity of NGC~6866}

We followed the procedure described in \cite{Karetal2003,Karetal2011} to measure
the photometric metallicity of the open cluster NGC~6866.
This method is mainly derived using F-G type main-sequence stars.
Therefore, we selected 17 out of 64 stars of the cluster
based on their colours ($0.3\leq (B-V)_0\leq 0.6$ mag) corresponding
to F0-G0 spectral type main-sequence stars \citep{Cox2000}.

We calculated the normalized ultraviolet (UV) excesses of the selected stars, which 
are defined as the differences between the de-reddened $(U-B)_0$ colour indices
of stars and the ones corresponding to the members of the Hyades cluster with the 
same de-reddened $(B-V)_0$ colour index, i.e. $\delta=(U-B)_{0,H}-(U-B)_{0,S}$. 
Here, the subscripts $H$ and $S$ refer to Hyades and star, respectively. Then we 
normalized the $\delta$ differences to the UV-excess at $(B-V)_{0}=0.6$ mag, 
i.e. $\delta_{0.6}$.

The $(U-B)_0$ vs $(B-V)_0$ TCD and the histogram of the normalized
$\delta_{0.6}$ UV excesses of the selected 17 main-sequence stars of
NGC~6866 are shown in Fig. 12. A Gaussian fit to the resulting histogram
allows us to calculate the normalized UV excess as $\delta_{0.6}=0.030\pm 0.005$
mag, where the uncertainty is given as the statistical uncertainty of the peak
of the Gaussian. In order to estimate the metallicity ($[Fe/H]$) of the cluster,
this mode value was evaluated in the following equation of \cite{Karetal2011}:

\begin{eqnarray}
[Fe/H]=-14.316(1.919)\delta_{0.6}^2-3.557(0.285)\delta_{0.6}+0.105(0.039).
\end{eqnarray}
The metallicity corresponding to the mode value for the $\delta_{0.6}$ distribution 
was calculated as $[Fe/H]= -0.013\pm 0.002$ dex.

The theoretical stellar evolutionary isochrones use the mass fraction $Z$ of all
elements heavier than helium. Thus, we used the following relation to transform
the $[Fe/H]$ metallicities obtained from the LAMOST spectra and the photometry
to the mass fraction $Z$ \citep{Mowlavietal2012}:

\begin{eqnarray}
Z=\frac{0.013}{0.04+10^{-[Fe/H]}}.
\end{eqnarray}

Hence, we calculated $Z=0.010$ and $Z=0.012$ from the $[Fe/H]$ metallicities
obtained from the LAMOST spectra and the photometry, respectively. Since these abundances
are very close to each other and to the solar value, which is given as $Z=0.0154$
by \cite{Bresetal2012}, we prefer to use the solar abundances in the determination of
the astrophysical parameters of the cluster. 

\subsection{Distance modulus and age of NGC~6866}

The reddening, metallicity, distance modulus and age of a cluster can be
simultaneously determined by fitting the theoretical stellar evolutionary
isochrones to the observed CMDs. In our case, we measured metallicity and reddening
of the cluster using reliable traditional methods, as noted above. Then, in
order to derive its distance modula and age simultaneously, we fitted the CMDs
of NGC~6866 with the theoretical isochrones provided by the PARSEC synthetic
stellar library \citep{Bresetal2012}, which was recently updated
\citep{Tangetal2014,Chenetal2014}. Since we keep the
metallicity and reddening of the cluster as constants during the fitting process,
it is expected that the degeneracy/indeterminacy of the parameters will be less
than that in the statistical solutions with four free astrophysical parameters.
In Fig. 13, the best fitted theoretical isochrones given by \cite{Bresetal2012} 
for $Z=0.0154$ and $t=800-850~\rm{Myr}$ are overplotted in the CMDs. The estimated 
astrophysical parameters of NGC~6866 obtained from the best fits to the CMDs are 
given in Table 7. Errors of the parameters are derived by visually shifting 
the theoretical isochrones to include all the main-sequence stars in the observed 
CMDs.

\begin{figure}
\begin{center}
\includegraphics[scale=0.7, angle=0]{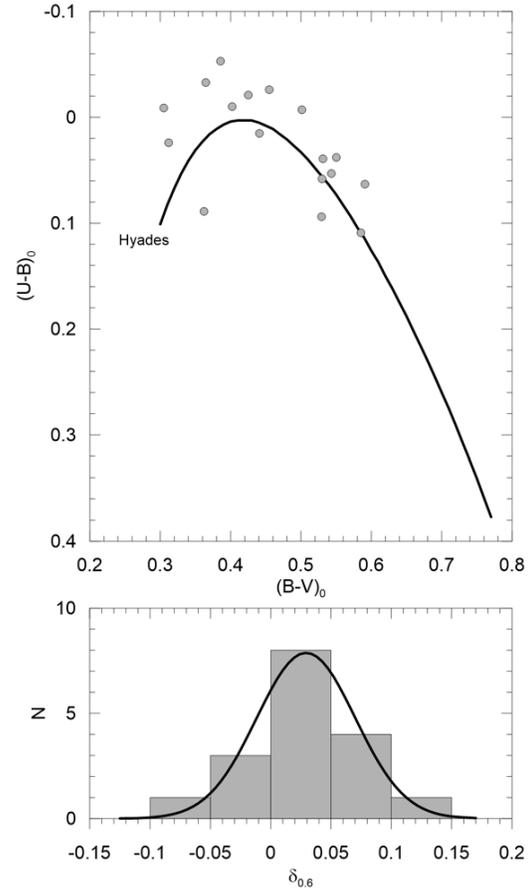}
\caption[] {\small The $(U-B)_{0}$ vs $(B-V)_{0}$ TCD ({\it upper
    panel}) and the histogram ({\it lower panel}) for the reduced
  UV-excess for 17 main-sequence stars used for the metallicity
  estimation of NGC 6866. The solid line in the upper panel represents
  the main-sequence of Hyades cluster while the one in the lower panel
  shows the Gaussian fit of the histogram.}
\end{center}
\end{figure}

\begin{table}
\setlength{\tabcolsep}{3pt}
\caption{Colour excesses, metallicities ($Z$), distance moduli ($\mu$), distances ($d$) and ages ($t$)
  estimated using four CMDs. The mean values are given in the last line.}
\begin{center}
{\tiny
\begin{tabular}{lccccc}
\hline
CMD          & Colour Excess & $Z$   & $\mu$ &    $d$  &  $t$    \\
             &      (mag)    &       & (mag) &    (pc)   & (Myr)    \\
\hline
$V$ vs $U-B$ & $E(U-B)=0.054\pm0.036$ & $0.0154$ & $10.75\pm 0.11$ &  $1271\pm 83$ & $850\pm50$ \\
$V$ vs $B-V$ & $E(B-V)=0.074\pm0.050$ & $0.0154$ & $10.53\pm 0.11$ &  $1148\pm 75$ & $800\pm50$ \\
$V$ vs $V-R$ & $E(V-R)=0.048\pm0.050$ & $0.0154$ & $10.48\pm 0.10$ &  $1122\pm 68$ & $800\pm50$ \\
$V$ vs $R-I$ & $E(R-I)=0.044\pm0.050$ & $0.0154$ & $10.65\pm 0.10$ &  $1214\pm 75$ & $800\pm50$ \\
\hline
\multicolumn{2}{r}{Mean} & $0.0154$ & $10.60\pm 0.10$ &  $1189\pm 75$ & $813\pm50$ \\
\hline
\end{tabular}
}
\end{center}
\end{table}

\begin{figure*}
\begin{center}
\includegraphics[scale=0.58, angle=0]{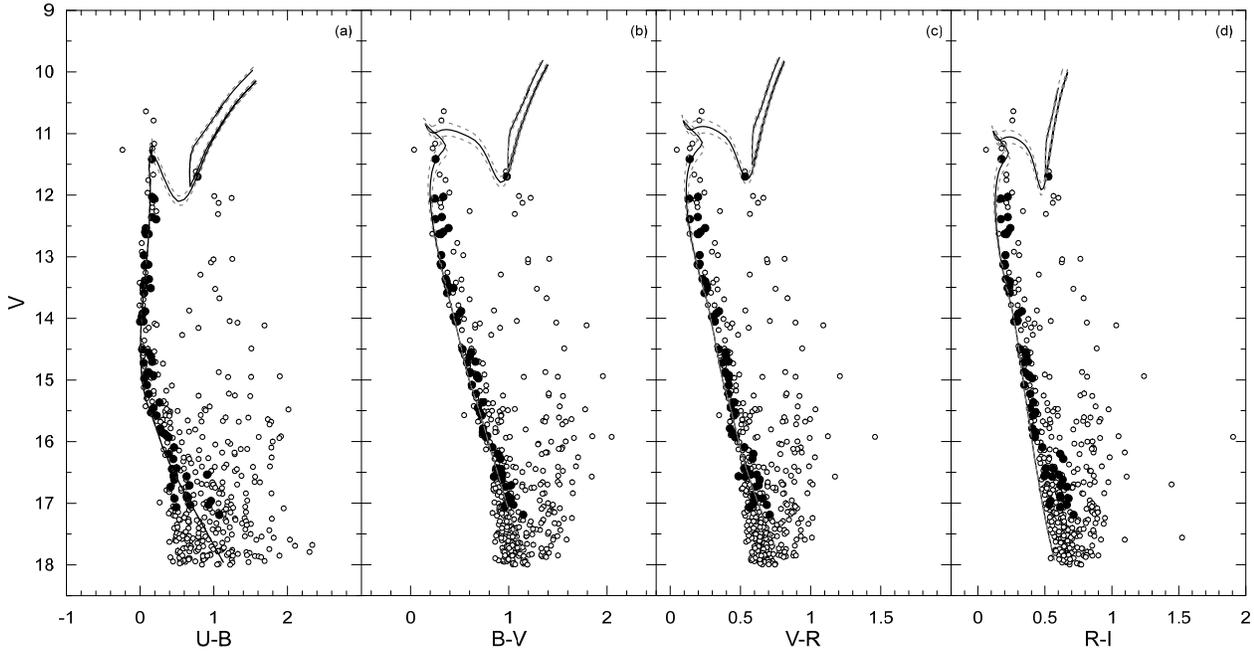}
\caption[] {\small Four CMDs for the stars located in a circle of 6
  arcmin radius from the center of NGC 6866. The most probable members
  of the cluster are indicated with black circles. These stars are
  fitted to the isochrone determined in this study (solid line). The
  dashed lines indicate the isochrones with estimated age
  plus/minus its error.}
\end{center}
\end{figure*}

\subsection{Galactic orbit of the cluster}

In order to estimate the parameters of the Galactic orbit of the cluster, we
followed the procedure described in \cite{Dinetal1999}, \cite{Coskunogluetal12}
and \cite{Bil12}. We first performed a test-particle integration in
a Milky Way potential which consists of a logarithmic halo, a Miyamoto-Nagai
potential to represent the Galactic disc and a Hernquist potential to model the bulge.
We then used values from \cite{Coskunogluetal11} for LSR corrections. 

We calculated Galactic orbits of the six cluster stars with a membership probability
larger than 50\% for which LAMOST data are given in Table 6. Median values
of the radial velocity and proper
motion components of these stars, and the mean distance of the cluster
were taken as the input parameters for the cluster's Galactic orbit
estimation: $V_{r} = 10.58$ km s$^{-1}$, $\mu_{\alpha}\cos{\delta}=-3.30$
and $\mu_{\delta}=-5.65$ mas yr$^{-1}$, and $d=1189$ pc, respectively.
The proper motion components were taken from \cite{Zachetal13}, while
the radial velocities of the cluster stars and their distances were
found in this study. Proper motions used in our study are almost
the same as used by \cite{Wuetal2009} ($\mu_{\alpha}\cos{\delta}=-3.33$
and $\mu_{\delta}=-5.03$ mas yr$^{-1}$) who also calculated the parameters
of the Galactic orbit of the cluster. 
Galactic orbits of the stars were determined within an
integration time of 3 Gyr in steps of 2 Myr. This integration time
corresponds to minimum 12 revolutions around the Galactic center so that
the averaged orbital parameters can be determined reliably. In order to
determine the Galactic orbit of the cluster, means of the orbital
parameters found for the stars were adopted as the orbital parameters of
the cluster.

In Fig. 14, representations of Galactic orbits calculated for the cluster
stars and the cluster itself are shown in the $X-Y$ and $X-Z$ planes. Here,
$X$, $Y$ and $Z$ are heliocentric Galactic coordinates directed towards the
Galactic centre, Galactic rotation and the north Galactic pole, respectively.
The cluster's apogalactic ($R_{max}$) and perigalactic ($R_{min}$) distances
were obtained as 9.78 and 7.76 kpc, respectively. The maximum vertical
distance from the Galactic plane is calculated as $Z_{max}= 160~{\rm pc}$.
When determining the eccentricity projected on to the Galactic plane, the
following formula was used: $e=(R_{max}-R_{min})/(R_{max}+R_{min})$. The
eccentricity of the orbit was calculated as $e=0.12$. This value shows
that the cluster is orbiting the Galaxy with a period of $P_{orb}= 156$ Myr
as expected for the objects in the solar neighborhood. 

\begin{figure}
\begin{center}
\includegraphics[trim=0cm 1cm 0cm 0cm, clip=true, scale=0.34]{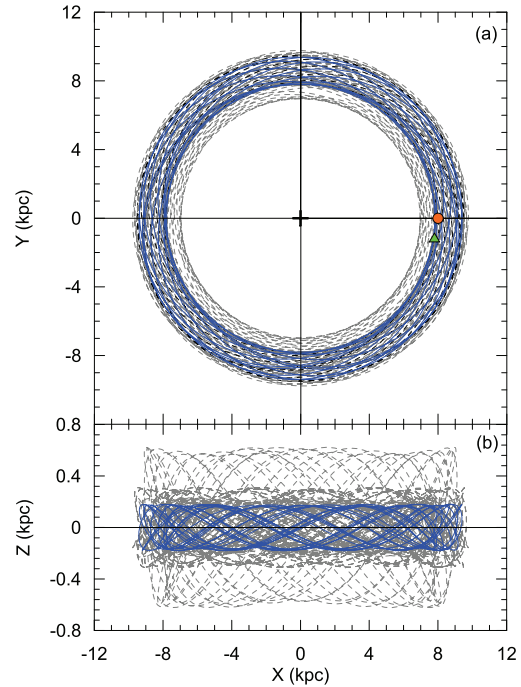}
\caption[] {\small The Galactic orbital motions (grey dashed lines) of 
the six cluster stars with $P>50\%$, for which LAMOST spectra are 
available, in the $X-Y$ (a) and $X-Z$ (b) planes. The cluster's mean 
orbit is indicated with a blue line. The black plus, red circle and 
green triangle symbols in panel (a) represent the Galactic centre, and current 
locations of the Sun and NGC 6866, respectively.}
\end{center}
\end{figure}

\subsection{Luminosity and mass functions of the cluster}
The luminosity function (LF) is defined as the relative number of
stars in the unit absolute magnitude range. Since the correction for
the non-member stars in a cluster field is very important in the
estimation of the luminosity function, we decided to demonstrate the
effect of non-member stars using the following procedure. First, we
selected the main-sequence stars with $V \leq 17$ mag located in a
circular field of 6 arcmin radius from the centre of the cluster. 
Note that the stars fainter than $V\approx 17$ mag have no accurate 
astrometric data in the UCAC4 catalogue \citep{Zachetal13}. Thus, 
the membership probabilities can not be estimated for the stars fainter 
than $V\approx 17$ mag. Then we removed the stars located out of the 
band-like region presented in Fig. 10. Additionally, we also 
selected the stars located near the turn-off point of the cluster and 
brighter than $V=12.75$ mag. This selection procedure resulted in 116 
stars with the membership probability $P > 0\%$. For the 
main-sequence stars, the apparent magnitude $V \leq 17$ corresponds to 
a mass range of $3.2 > M/M_\odot > 0.8$ for the cluster.  In order to 
demonstrate the effect of non-member field stars, the LFs of the cluster 
estimated for the stars with the membership probabilities of $P\geq 20\%$ 
($N=87$) and $P\geq 50\%$ ($N=64$) are displayed in Fig. 15. The 
LFs of NGC~6866 in Fig. 15 indicate that the LF of the cluster 
has a maximum value towards $M_{V}\approx 5.5$ mag.

The mass function (MF) denotes the relative number of stars in a unit range 
of mass centered on mass $M$. It represents the rate of star creation as a 
function of stellar mass. Theoretical models provided by the PARSEC synthetic
stellar library \citep{Bresetal2012} were used to convert the LFs to MFs for 
NGC~6866. Resulting MFs are presented in Fig. 16. The slope $x$ of mass function
can be derived by using the following linear relation: $\log (dN/dM)=-(1+x)\log(M)+C$,
where $dN$ represents the number of stars in a mass bin $dM$ with central mass of $M$,
and $C$ is a constant. The slopes of the MFs are found to be
$x=1.48 \pm 0.21$ and $x=1.35 \pm 0.08$ for the stars with the membership
probabilities of $P\geq 20\%$ and $P\geq 50\%$, respectively. Since these values
are in agreement in errors, we conclude that the effect of non-member stars can
be negligible. As the stars with the membership probability $P\geq 50\%$ constitute
a more pure sample for the cluster stars, we adopt the MF slope $x=1.35 \pm 0.08$ for
NGC~6866. This MF slope is in a perfect agreement with the value of 1.35
given by \cite{Salpe1955} for the stars in the solar neighbourhood.

\begin{figure}
\begin{center}
\includegraphics[trim=0cm 4.5cm 0cm 4cm, clip=true, scale=0.4]{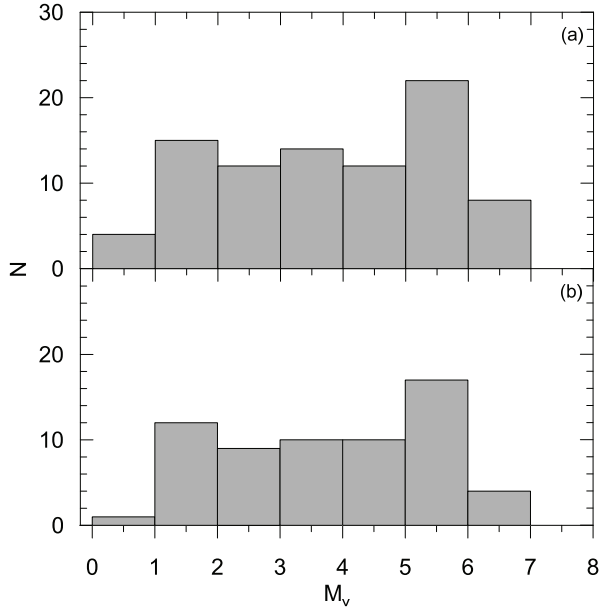}
\caption[] {\small The luminosity functions of NGC 6866 estimated
for the stars with the membership probability $P\geq 20\%$ (a) and
$P\geq 50\%$ (b).}
\end{center}
\end{figure}

\begin{figure}
\begin{center}
\includegraphics[trim=0cm 5cm 0cm 4cm, clip=true, scale=0.4]{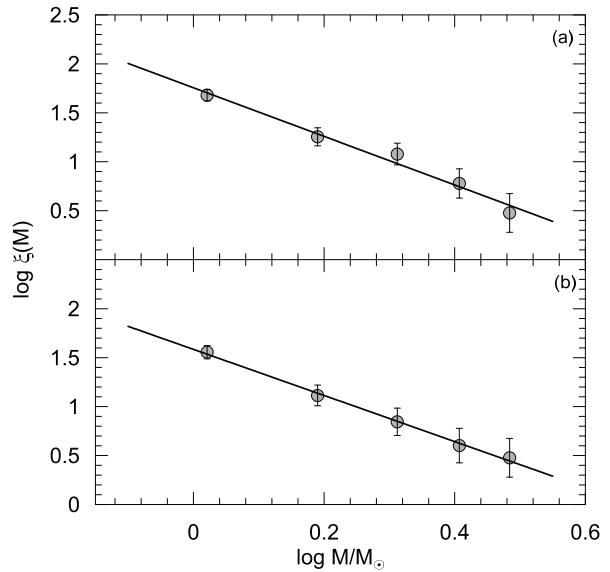}
\caption[] {\small The mass functions of NGC~6866 estimated for the
stars with the membership probability $P\geq 20\%$ (a) and
$P\geq 50\%$ (b).}
\end{center}
\end{figure}

\section{Discussion}

The reddening, metallicity, distance modulus and age of a cluster can
be simultaneously determined by fitting the theoretical stellar
evolutionary isochrones to the observed CMDs. However, astrophysical
parameters derived from the simultaneous solutions suffer from the
reddening-age degeneracy
\citep[cf.][]{Anders2004,Kingetal2005,Brid2008,deMeule2013}. Thus,
independent methods developed for the determination of the
astrophysical parameters are very useful to reduce the number of free
parameters. In this study, using such methods, the
reddening was inferred from the $U-B$ vs $B-V$ TCD of the cluster and
found as $E(B-V)=0.074\pm 0.050$ mag and the metallicities estimated
from the spectroscopic and photometric observations were found to be
roughly in agreement and assumed to be the solar value. Hence, keeping
these two parameters as constants, we derived the distance modula and
age of NGC~6866 by fitting its observed CMDs to the theoretical
isochrones as  $\mu = 10.60 \pm 0.10$ mag and $t = 813 \pm 50$ Myr,
respectively. With this method, we attempt to break in part the
reddening-age degeneracy.

Although the same metallicity values were adopted in the determination
of the astrophysical parameters of NGC~6866, a comparison of Tables 1
and 7 reveals that the reddening, the distance modula, and the distance
in this study were found to be smaller than those reported
in the previous studies. The determined age of the cluster is in
agreement with \cite{Gunetal2012} and \cite{Janes2014}. Since our
photometric data is in good agreement with those of \cite{Joshietal2012} 
and \cite{Janes2014}, we conclude that the main reason for the disagreement 
of the results can be due to the chosen isochrone and age determination 
difference. It would be confident to conclude that our results are reliable 
since a more comprehensive approach which tries to minimize degeneracies 
among parameters is used in this study.

There is only one star with $P>50\%$ in the RC region of the CMDs of the 
cluster. The apparent magnitude and the colour index for this star are 
$V= 11.699$ and $B-V= 0.979$ mag, respectively. We adopt 
$E(B-V)=0.074\pm 0.050$ (see Section 4.1) and $M_V=1.0\pm0.2$ mag 
\citep{Biletal2013b} as the intrinsic colour excess and absolute 
magnitude of this RC star, respectively. Using the Pogson's relation, 
the distance of the RC star of the cluster is calculated as 
$d=1241 \pm 110$ pc, in agreement with the mean distance given in 
Table 7, $d= 1189 \pm 75$ pc.

Using the most likely members of the cluster for which the radial
  velocities are taken from the LAMOST database, we estimated the
  Galactic orbit of NGC~6866. The eccentricity projected on to the
  Galactic plane for its orbit is calculated to be $e=0.12$, which
  implies that the cluster is located in the thin-disc component of
  the Galaxy. Galactic orbital parameters of the cluster were also
  calculated by \cite{Wuetal2009} using the mean radial velocity given
  in \cite{Meretal2008}. They calculated the eccentricity of the orbit
  of the cluster to be $e=0.08$ and the maximum vertical distance from
  the Galactic plane $Z_{max}= 220~{\rm pc}$. A comparison shows that
  the eccentricity found in our study is about 50\% larger than that
  calculated in \cite{Wuetal2009} probably since $Z_{max}$ value in
  this study is about 30\% smaller than their ones. 

  The luminosity function of NGC~6866 has a maximum value near
  $M_{V}\approx 5.5$ mag. The derived slope of the mass function for
  the cluster is $x=1.35 \pm 0.08$ which is in a good agreement with
  the value of 1.35 given by \cite{Salpe1955} for the stars in the
  solar neighbourhood.

\section{Acknowledgments}

Authors are grateful to the anonymous referee for his/her considerable
contributions to improve the paper. This work has been supported in part by 
the Scientific and Technological Research Council (T\"UB\.ITAK) 113F201 and 113F270.
Part of this work was supported by the Research Fund of the University of Istanbul,
Project Numbers: 39170 and 39742. We thank to T\"UB\.ITAK National Observatory for 
a partial support in using T100 telescope with project number 12BT100-324. Guoshoujing
Telescope (the Large Sky Area Multi-Object Fiber Spectroscopic Telescope,
LAMOST) is a National Major Scientific Project which is built by the Chinese Academy
of Sciences, funded by the National Development and Reform Commission, and operated
and managed by the National Astronomical Observatories, Chinese Academy of Sciences.
Funding for the project has been provided by the National Development and Reform 
Commission. LAMOST is operated and managed by the National Astronomical Observatories, 
Chinese Academy of Sciences. EP acknowledge support by the SoMoPro II programme 
(3SGA5916). It was also supported by the grants GP14-26115P, 7AMB14AT015, the financial 
contributions of the Austrian Agency for International Cooperation in Education and
Research (BG-03/2013 and CZ-09/2014). This research has made use of the
WEBDA, SIMBAD, and NASA\rq s Astrophysics Data System Bibliographic Services.

\end{document}